\documentclass[article,prl]{revtex4}
\usepackage{amssymb}
\usepackage[T1]{fontenc}
\usepackage[latin9]{inputenc}
\usepackage{color}
\usepackage{amsmath}
\usepackage{graphicx}
\usepackage{esint}
\usepackage{bm}
\usepackage{braket}
\usepackage{bbm}

\makeatletter
\@ifundefined{textcolor}{}
{ \definecolor{BLACK}{gray}{0}
 \definecolor{WHITE}{gray}{1}
 \definecolor{RED}{rgb}{1,0,0}
 \definecolor{GREEN}{rgb}{0,1,0}
 \definecolor{BLUE}{rgb}{0,0,1}
 \definecolor{CYAN}{cmyk}{1,0,0,0}
 \definecolor{MAGENTA}{cmyk}{0,1,0,0}
 \definecolor{YELLOW}{cmyk}{0,0,1,0}
 }

\begin{document}

\title{Possible Vulnerability of Bell-Clauser-Horne-Shimony-Holt Tests used for Quantum Certification}

\author{F. De Zela}
\affiliation{Departamento de Ciencias, Secci\'{o}n
F\'{i}sica, Pontificia Universidad Cat\'{o}lica del Per\'{u}, Lima 15088, Peru.}

\begin{abstract}
A hidden variables (HVs) model is reported, which reproduces quantum predictions for Bell-Clauser-Horne-Shimony-Holt (Bell-CHSH) tests. The existence of such a model poses some limitations to quantum certifications that rely on Bell-CHSH inequality violations. The reported model does not prove wrong Bell's theorem. The latter assumes the factorability of the probability density $p_{AB}$, which rules the stochastic behavior of the HVs. The reported HVs model is based on an extended form of $p_{AB}$, which is suggested by Lebesgue's decomposition theorem for bounded functions. The considered $p_{AB}$ complies with locality and realism, and also with measurement independence, parameter independence and outcome independence. 
\end{abstract}

\maketitle

\section{Introduction}

Once developed and produced at industrial levels, both quantum computers and devices enabling quantum communication will require massive usage of entangled states. These states guarantee the availability of non-classical input-output correlations, a basic resource for reaching quantum advantage. Bell tests offer themselves as a handy tool for certifying entanglement. Among all Bell tests, the most practical and experimentally studied one is the Bell-Clauser-Horne-Shimony-Holt (Bell-CHSH) test. A quantum device that produces entangled states should be able to violate the Bell-CHSH inequality: $|S|\leq 2$. Here, $S$ is a parameter which is constructed out of correlations between measurement outcomes at two distant sites. The Bell-CHSH inequality is the logical consequence of some assumptions that are fulfilled by all classical resources, i.e., those which should be outperformed by quantum resources. All Bell tests that have been performed up to now required making some additional assumptions. These give rise to loopholes in the logical consequences of the original assumptions. For this reason, Bell-CHSH inequality violations cannot fully guarantee the availability of non-classical correlations. In other words, Bell-CHSH inequality violations could be obtained, at least in principle, with classical resources. Depending on the loophole, the imagined working of these classical resources can be more or less contrived or even unrealistic. Loopholes giving rise to unrealistic schemes to produce Bell-CHSH inequality violations can be dismissed, in practice, as serious threats
for a reliable certification of quantum resources. The remaining, in principle exploitable loopholes, have required considerable technical efforts to be closed. A few years ago, three different groups almost simultaneously announced to have closed these loopholes \cite{hensen1,giustina,shalm}. In contrast to the rather bold and general way in which the announcements were disseminated, the reported achievements were more precisely and cautiously framed, in each case. It was namely noted that, ``strictly speaking, no Bell experiment can exclude all conceivable local-realist theories'' \cite{hensen1}, that the reported experiments provided just ``the strongest support to date for the viewpoint that local realism is untenable'' \cite{giustina}, and that the announced test \cite{shalm} was loophole free, insofar as only a minimal set of assumptions was made, while ``It is impossible, even in principle, to eliminate a form of these assumptions in any Bell test'' \cite{shalm}. In regard to foundational issues, there seems to be a consensus that locality and realism have not been satisfactorily banned from physics yet \cite{valdenebro,ringbauer,hensen2,santos2022}. For that to be the case, more compelling evidence must be provided. Hence, increasingly sophisticated tests are currently under planning \cite{belenchia2022,mohageg2022}.

Recently, a ``loophole-free'' Bell-CHSH test was conducted using superconducting qubits \cite{storz2023}, a promising resource for building large-scale quantum computers and communication devices. Various loopholes were simultaneously closed, namely the locality loophole, which appears whenever two distant actions are not guaranteed to be spacelike separated; the fair-sampling loophole, which shows up when measurements are biased; and the memory loophole, which occurs when measurement devices retain information from already performed measurements, and this can influence subsequent ones. On closing all these loopholes, a major step forward was made towards convincingly exhibiting nonlocal correlations and implementing secure, device-independent quantum communication protocols. However, a remaining, important assumption had not been addressed. The parameter $S$ entering the Bell-CHSH inequality involves four correlations. Each entangled pair can be used to measure only one of these correlations. This is because the observables involved in the correlations do not commute with each other, and therefore cannot be simultaneously measured. Thus, independent experimental runs must be performed on identical copies of the same entangled state. This gives rise to another loophole, which stems from making the additional assumption of counterfactual definiteness. That is, one must assume that it is possible to meaningfully refer to results of measurements that have not been performed \cite{luis2025,genovese2025,comments}. Such an additional assumption can be avoided by performing all measurements on one and the same entangled pair. This has been achieved only recently, by resorting to weak measurement techniques \cite{virzi2024,atzori2024}. One can then not only extract the Bell-CHSH parameter $S$ from each entangled pair, but also avoid destroying entanglement as a consequence of measurement. This opens the way to certify entanglement on resources that remain available for performing quantum tasks. It is of course a major challenge to apply weak measurement techniques when closing the other, aforementioned loopholes.

While it is unlikely that all possible loopholes will ever be closed, the previously mentioned ones must be closed before quantum technologies are deemed sufficiently reliable to be massively deployed. The present work addresses a possible obstacle that quantum certification could face in the future. This obstacle stems from the limited scope that Bell correlations actually have. It will be shown that Bell correlations are not the only ones that comply with locality and realism. 
This could suggest how to produce Bell-CHSH inequality violations without quantum entanglement, thereby exhibiting possible vulnerabilities of Bell-CHSH tests that certify the availability of quantum resources.

\section{Bell's model and the Bell-CHSH inequality}

The vulnerability of Bell-CHSH tests becomes particularly clear in a recent work \cite{genovese2025}. This work addressed the previously mentioned Bell-CHSH tests that were conducted using weak measurement techniques \cite{virzi2024,atzori2024}. According to Ref.~\cite{genovese2025}, the hypotheses leading to the Bell-CHSH inequality are three: (1) \emph{Realism}, i.e., the assumption that objects have physical properties that exist independently of being measured. Some of these properties are linked to hidden variables (HVs), whose stochastic realization is ruled by  some probability density. HVs are supposed to restore determinism but remain inaccessible to us, giving rise to an apparent randomness of some physical phenomena. (2) \emph{Independence of measurements} that are carried out at spacelike separated locations. (3) \emph{Bell locality}, also called local factorizability or local causality \cite{vieira2025}, which states that the joint probability density of the correlations submitted to test is a product of two probability densities. Below, we give the precise mathematical formulation of this last assumption. For now, we just notice that it is this third assumption the one that gives rise to the limitations of quantum certifications via Bell-CHSH tests. To see this, let us remind how the Bell-CHSH inequality follows from the above, three hypotheses.   

\begin{figure}[h!!]
\centering
\includegraphics[width=0.7\linewidth]{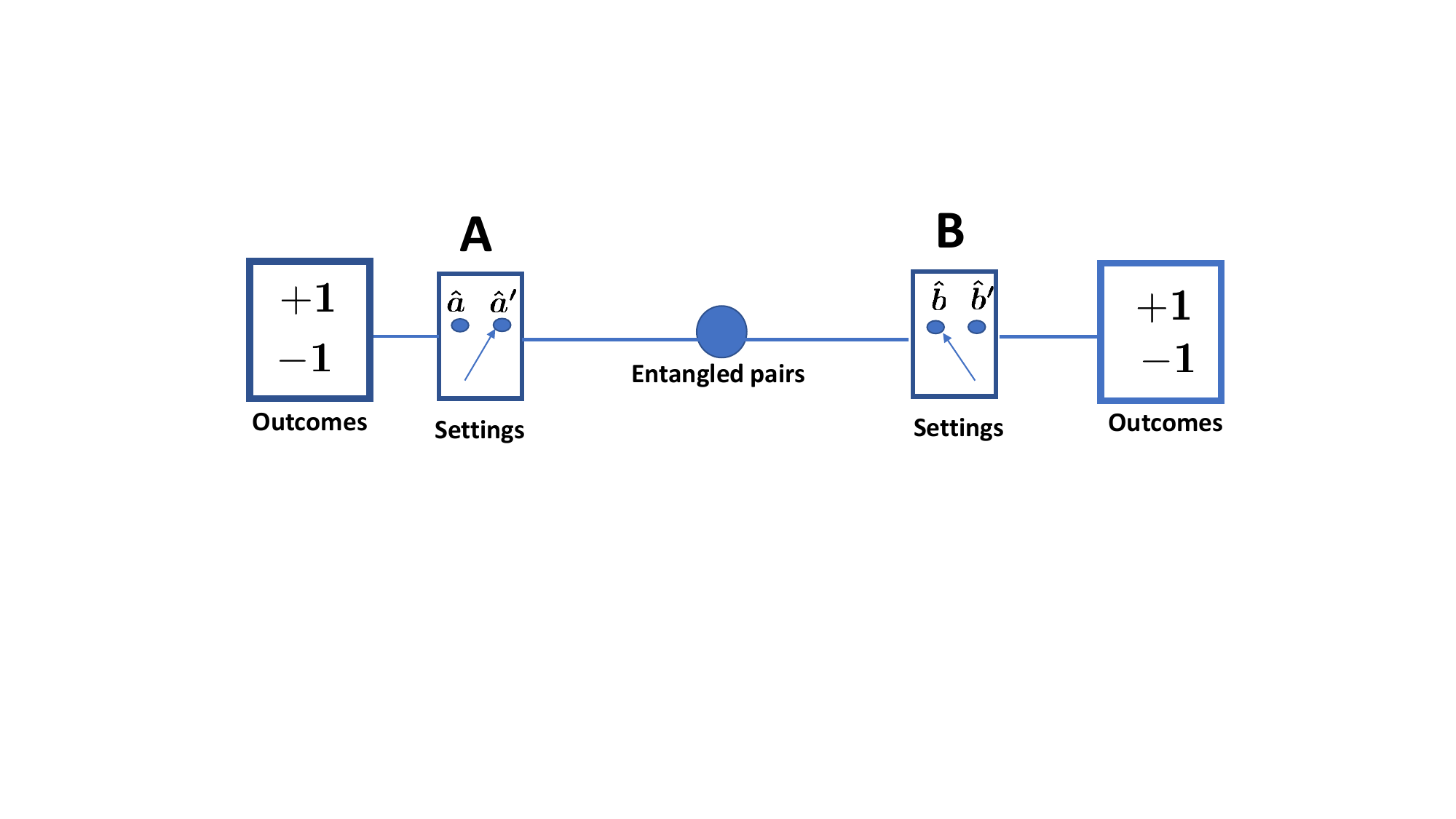}
\caption{Bell-CHSH experiment: Entangled pairs of spin-1/2 particles are produced by a source. One particle is sent to site $A$ and the other to site $B$. On each site, SG-magnets have a binary outcome ($\pm 1$). Each magnet can be oriented along two directions, $\boldsymbol{\hat{a}}$, $\boldsymbol{\hat{a}'}$ at $A$, and
$\boldsymbol{\hat{b}}$, $\boldsymbol{\hat{b}'}$ at $B$. The raw data of the experiment consist of simultaneous detections at $A$ and $B$.}
\label{fig1}
\end{figure}

The Bell-CHSH inequality involves Bell-type correlations between measurements of spin observables, $\hat{A}$ and $\hat{B}$, that are performed by two parties, Alice and Bob, at two distant sites, $A$ and $B$, respectively. Instead of spin observables, other two-state observables can be used, e.g., polarized states of light or superconducting qubits. The spin measurements at site $A$ are made by aligning a Stern-Gerlach (SG) magnet along one of two directions, $\boldsymbol{\hat{a}}$ and  $\boldsymbol{\hat{a}'}$, and similarly at $B$, along directions
$\boldsymbol{\hat{b}}$ and $\boldsymbol{\hat{b}'}$. The particles, on which Alice and Bob perform their measurements, are supplied by a source that produces pairs of spin-1/2 particles in an entangled state.
There are two detectors at the output of each SG magnet (see Fig.~(\ref{fig1})). The raw experimental data is the number of coincident detections during some fixed interval of time.
In a HVs model, the observables $\hat{A}$ and $\hat{B}$ are assumed to be given by scalar functions of the HVs $\lambda \in \Lambda$. For example, $A_{\boldsymbol{\hat{a}}}(\lambda)=\pm 1$ represents the binary output of Alice's SG magnet when it is oriented along $\boldsymbol{\hat{a}}$, this output being assumed to be fully determined by the HVs $\lambda$. $B_{\boldsymbol{\hat{b}}}(\lambda)=\pm 1$ is similarly defined. The probability
$P_{\boldsymbol{\hat{a}},\boldsymbol{\hat{b}}}(\alpha,\beta)$ for obtaining outputs $\alpha$ and $\beta$, given the settings $\boldsymbol{\hat{a}}$ and $\boldsymbol{\hat{b}}$, respectively, is theoretically modeled by 
\begin{equation}
 P_{\boldsymbol{\hat{a}},\boldsymbol{\hat{b}}}(\alpha,\beta)\equiv\int_{\Lambda} p(\alpha,\beta|\boldsymbol{\hat{a}},\boldsymbol{\hat{b}},\lambda)\rho(\lambda) d\lambda
 =\int_{\Lambda} p(\alpha|\boldsymbol{\hat{a}},\lambda)\,p(\beta|\boldsymbol{\hat{b}},\lambda)\rho(\lambda) d\lambda. \label{connection1}
\end{equation}
Here, $p(\alpha,\beta|\boldsymbol{\hat{a}},\boldsymbol{\hat{b}},\lambda)$ is the conditional probability density for Alice and Bob obtaining the values $\alpha$ and $\beta$ in their spin measurements, provided these measurements were performed with their SG magnets oriented along
$\boldsymbol{\hat{a}}$ and $\boldsymbol{\hat{b}}$, respectively, while the HVs had the value $\lambda$. The probability distribution of the HVs is ruled by the probability density $\rho(\lambda)\geq 0$, with $\int \rho(\lambda)d\lambda=1$. Likewise,
$p(\alpha|\boldsymbol{\hat{a}},\lambda)$ denotes the (conditional) probability density for Alice's spin measurement to yield $\alpha$, if her SG magnet was oriented along $\boldsymbol{\hat{a}}$ and the HVs had the value $\lambda$. Similarly for $p(\beta|\boldsymbol{\hat{b}},\lambda)$. In Eq.~(\ref{connection1}), the third hypothesis, \emph{Bell locality}, was applied. In mathematical terms, this third hypothesis states that the probability density for $\alpha$ and $\beta$ factorizes as follows:
\begin{equation}\label{factor}
 p(\alpha,\beta|\boldsymbol{\hat{a}},\boldsymbol{\hat{b}},\lambda)=
 p(\alpha|\boldsymbol{\hat{a}},\lambda)\,p(\beta|\boldsymbol{\hat{b}},\lambda).
\end{equation}
This decomposition ``represents a precise condition for locality in the context of Bell experiments''
\cite{brunner2014}. Below, we will discuss in detail the validity of such an assertion. Here, it should suffice to stress that the decomposition in Eq.~(\ref{factor}) involves probability densities, instead of probabilities.

The probability densities in Eq.~(\ref{connection1}) can be written as $p(\alpha|\boldsymbol{\hat{a}},\lambda)=(1+\alpha A_{\boldsymbol{\hat{a}}}(\lambda))/2$ and $p(\beta|\boldsymbol{\hat{b}},\lambda)=(1+\beta B_{\boldsymbol{\hat{b}}}(\lambda))/2$, so that
\begin{equation}\label{connection}
 P_{\boldsymbol{\hat{a}},\boldsymbol{\hat{b}}}(\alpha,\beta)=\int_{\Lambda} \left(\frac{1+\alpha A_{\boldsymbol{\hat{a}}}(\lambda)}{2}\right)\left(\frac{1+\beta B_{\boldsymbol{\hat{b}}}(\lambda)}{2}\right)\rho(\lambda) d\lambda.
\end{equation}
Measurable correlations between observables $\hat{A}$ and $\hat{B}$ can be defined as $\langle \hat{A}\hat{B} \rangle\equiv\sum_{\alpha,\beta} \alpha \beta P_{\boldsymbol{\hat{a}},\boldsymbol{\hat{b}}}(\alpha,\beta)$.
From Eq.~(\ref{connection}), we then obtain
\begin{equation}\label{expcor2}
\langle \hat{A}\hat{B} \rangle=\sum_{\alpha,\beta} \alpha \beta P_{\boldsymbol{\hat{a}},\boldsymbol{\hat{b}}}(\alpha,\beta) = \int_{\Lambda} A_{\boldsymbol{\hat{a}}}(\lambda)B_{\boldsymbol{\hat{b}}}(\lambda)\rho(\lambda)d\lambda.
\end{equation}
The above correlations are constrained by the Bell-CHSH inequality. To show this, a CHSH measure is first defined as follows:
\begin{equation}\label{bm1}
 S_{\text{CHSH}}=\langle \hat{A}\hat{B} \rangle+\langle \hat{A}\hat{B}^{\prime} \rangle +\langle \hat{A}^{\prime}\hat{B} \rangle -\langle \hat{A}^{\prime}\hat{B}^{\prime} \rangle.
\end{equation}
Here, we have written $\hat{A}$ and $\hat{A}^{\prime}$ when referring to the settings $\boldsymbol{\hat{a}}$ and $\boldsymbol{\hat{a}'}$, respectively, and similarly for $\hat{B}$ and $\hat{B}^{\prime}$.
The structure of $S_{\text{CHSH}}$ comes from considering the function
$f(\lambda)=(A(\lambda)+A^{\prime}(\lambda))B(\lambda)+(A(\lambda)-A^{\prime}(\lambda))B^{\prime}(\lambda)$,
which is the sum of two terms, one of which is $\pm 2$ and the other zero for each $\lambda$, whereupon $f(\lambda)=\pm 2$. From this and
$|\int_{\Lambda} f(\lambda)d\lambda| \leq \int_{\Lambda} |f(\lambda)|d\lambda=2$, one gets the Bell-CHSH inequality:
\begin{equation}\label{s}
  |S_{\text{CHSH}}|\leq 2.
\end{equation}

It is worth mentioning that Bell originally \emph{defined} the correlations $\langle \hat{A}\hat{B} \rangle=\int A_{\boldsymbol{\hat{a}}}(\lambda)B_{\boldsymbol{\hat{b}}}(\lambda)\rho(\lambda)d\lambda$ \cite{bell1964}. Bell took $A_{\boldsymbol{\hat{a}}}(\lambda)$ and $B_{\boldsymbol{\hat{b}}}(\lambda)$ to be the results of measuring the quantum observables $\boldsymbol{\hat{a}} \cdot \boldsymbol{\sigma}^{A}$ and $\boldsymbol{\hat{b}} \cdot \boldsymbol{\sigma}^{B}$, respectively, where $\boldsymbol{\sigma}=(\sigma_x,\sigma_y,\sigma_z)$ is the triple of Pauli matrices, while $\boldsymbol{\hat{a}}$ and $\boldsymbol{\hat{b}}$ are unit vectors. Hence, $A_{\boldsymbol{\hat{a}}}(\lambda)$ and $B_{\boldsymbol{\hat{b}}}(\lambda)$ must have the values $\pm 1$. Bell's expression for $\langle \hat{A}\hat{B}\rangle$ follows from assuming that the quantum expectation value 
$\langle\boldsymbol{\hat{a}} \cdot \boldsymbol{\sigma}^{A}\otimes\boldsymbol{\hat{b}} \cdot \boldsymbol{\sigma}^{B}\rangle$ is given by 
$\langle \hat{A}\hat{B} \rangle=\int (A_{\boldsymbol{\hat{a}}}B_{\boldsymbol{\hat{b}}})(\lambda)\rho(\lambda)d\lambda $ in a HV model that ``completes'' quantum mechanics (QM). Locality is subsequently enforced by setting $(A_{\boldsymbol{\hat{a}}}B_{\boldsymbol{\hat{b}}})(\lambda)=A_{\boldsymbol{\hat{a}}}(\lambda)B_{\boldsymbol{\hat{b}}}(\lambda)$. The derivation of Eq.~(\ref{expcor2}) from the probabilities $P_{\boldsymbol{\hat{a}},\boldsymbol{\hat{b}}}(\alpha,\beta)$
in Eq.~(\ref{connection}) is an elaboration of Bell's argumentation that has been presented by several authors (see, e.g., \cite{aspect2002,dehlinger2002}). The enforcement of locality, either by setting $(A_{\boldsymbol{\hat{a}}}B_{\boldsymbol{\hat{b}}})(\lambda)=A_{\boldsymbol{\hat{a}}}(\lambda)B_{\boldsymbol{\hat{b}}}(\lambda)$ or else by Eq.~(\ref{factor}), represents a special case, as has been pointed out before. For instance, Branciard et al. \cite{branciard} refer to Eq.~(\ref{factor}) as ``the simplest choice'' for enforcing Bell's locality assumption and set out to explore other choices, in particular those leading to Legget's inequalities \cite{groblacher,leggett,dilorenzo2012}. 

\section{Hidden variables model that violates the Bell-CHSH inequality}

\subsection{Preliminary remarks}

As is well known, QM predicts the violation of the Bell-CHSH inequality, Eq.~(\ref{s}). This has led to take violations of Eq.~(\ref{s}) as a possible means for quantum certification. As any certification, quantum certifications can be issued only with the proviso that some assumptions have been met. In what follows, we focus on two assumptions that are essential to derive the Bell-CHSH inequality. These assumptions strongly restrict the class of HVs models that can be falsified by experimental violations of the Bell-CHSH inequality. In mathematical terms, the assumptions read
\begin{eqnarray}
  &1.& \quad 0 \leq  p(\alpha|\boldsymbol{\hat{a}},\lambda),\,p(\beta|\boldsymbol{\hat{b}},\lambda)\leq  1, \label{c1} \\
  &2.& \quad p(\alpha,\beta|\boldsymbol{\hat{a}},\boldsymbol{\hat{b}},\lambda)= p(\alpha|\boldsymbol{\hat{a}},\lambda)\,p(\beta|\boldsymbol{\hat{b}},\lambda). \label{c2}
\end{eqnarray}
Assumption 1 says that the probability densities must be restricted to take values in the interval $[0,1]$. This is indeed the case with the probability densities in Eq.~(\ref{connection1}), i.e., $p(\alpha|\boldsymbol{\hat{a}},\lambda)=(1+\alpha A_{\boldsymbol{\hat{a}}}(\lambda))/2$ and $p(\beta|\boldsymbol{\hat{b}},\lambda)=(1+\beta B_{\boldsymbol{\hat{b}}}(\lambda))/2$, which have the values 0 or 1 for each $\lambda$ \cite{aspect2002,dehlinger2002}. In alternative derivations, the probability densities may vary within the whole interval $[0,1]$ \cite{clauser1978,clauser1974}. The inequalities in assumption 1 allegedly ``hold if the probabilities are sensible'' \cite{clauser1974}. It thus seems that assumption 1 was made on taking probability \emph{densities} as if they were probabilities. Irrespective of its motivation, the fact is that assumption 1 severely restricts the set of HV models for which quantum certification via Bell-CHSH inequality violations applies. This set excludes models based on probability densities which are often used in physics. A prominent example is the Gaussian density $p(\lambda)=\exp\left(-(\lambda-\mu)^2/2\sigma^2\right)/\sqrt{2\pi\sigma^2}$, which has values $p(\lambda)\gg 1$ whenever $\sigma \ll 1$.  

Assumption 2 -- commonly called ``local factorizability'' or ``local causality'' -- has been more amply discussed than assumption 1. For instance, Clauser and Horne argued that the factored form of
$p(\alpha,\beta|\boldsymbol{\hat{a}},\boldsymbol{\hat{b}},\lambda)$ ``is a natural expression of [...] the common-sense view that there is no action at the distance'' \cite{clauser1974}. Thus, the \emph{probability} \cite{clauser1974} for obtaining the result $\alpha$ with setting $\boldsymbol{\hat{a}}$ at site $A$ does not depend on setting $\boldsymbol{\hat{b}}$ at site $B$ nor on result $\beta$, and viceversa. Bell argued similarly \cite{bell1964}: ``The vital assumption [2] is that the result B for particle 2 does not depend on the setting $\boldsymbol{\hat{a}}$, of the magnet for particle 1, nor $A$ on $\boldsymbol{\hat{b}}$''. In citation ``[2]'' of his text, Bell quoted Einstein, who said: ``But on one supposition we should, in my opinion, absolutely hold fast: the real factual situation of the system $S_2$ is independent of what is done with system $S_1$, which is spatially separated from the former''.

The problem with assumption 2 is that all arguments, which are invoked to justify the factored form of $p(\alpha,\beta|\boldsymbol{\hat{a}},\boldsymbol{\hat{b}},\lambda)$, should equally well apply to the actual probability $P(\alpha,\beta|\boldsymbol{\hat{a}},\boldsymbol{\hat{b}})$. However, on assuming that $P(\alpha,\beta|\boldsymbol{\hat{a}},\boldsymbol{\hat{b}})= P(\alpha|\boldsymbol{\hat{a}})\,P(\beta|\boldsymbol{\hat{b}})$, we are led to 
\begin{equation}\label{expcor2a}
\langle \hat{A}\hat{B} \rangle=\sum_{\alpha,\beta} \alpha \beta P(\alpha,\beta|\boldsymbol{\hat{a}},\boldsymbol{\hat{b}})=\sum_{\alpha} \alpha P(\alpha|\boldsymbol{\hat{a}})\, \sum_{\beta} \beta P(\beta|\boldsymbol{\hat{b}}) =\langle \hat{A}\rangle\langle \hat{B} \rangle,
\end{equation} 
which are the correlations that correspond to independent events. This would make the Bell-CHSH inequality pointless. Indeed, setting 
$\langle \hat{A}\hat{B} \rangle=\langle\hat{A}\rangle\langle \hat{B} \rangle$, etc., in $S_{\text{CHSH}}$, we get
\begin{equation}\label{s2}
 S_{\text{CHSH}}=\langle \hat{A}\rangle\langle \hat{B} \rangle+\langle \hat{A}\rangle\langle \hat{B}^{\prime} \rangle +\langle \hat{A}^{\prime}\rangle\langle \hat{B} \rangle -\langle \hat{A}^{\prime}\rangle\langle \hat{B}^{\prime} \rangle,
\end{equation}
and the corresponding inequality $|S_{\text{CHSH}}|\leq 2$ would be always satisfied in Bell tests, because 
$\langle\boldsymbol{\hat{a}} \cdot \boldsymbol{\sigma}^{A}\rangle=0=\langle\boldsymbol{\hat{b}} \cdot \boldsymbol{\sigma}^{B}\rangle$, for fully entangled states.   

It should then be clear that Bell's factorability, Eq.~(\ref{c2}), is not imposed by mutual independence nor by its restricted variants, such as ``no-signaling'', ``locality'', ``relativistic causality'', etc. Any claim that Eq.~(\ref{c2}) necessarily follows from mutual independence -- or from restrictive variants thereof -- would apply not only for probability densities, but for probabilities as well. 
Hence, if we do \emph{not} assume a factorized form for the probability $P(\alpha,\beta|\boldsymbol{\hat{a}},\boldsymbol{\hat{b}})$, then we should take Eq.~(\ref{c2}) as being only a particular assumption among other possible ones that comply with local realism. One may, for instance, consider a probability density with the following structure:
\begin{equation}\label{factor3}
 p_{AB}(\alpha,\beta|\boldsymbol{\hat{a}},\boldsymbol{\hat{b}},\lambda)=
 f_A(\alpha|\boldsymbol{\hat{a}},\lambda)\,f_B(\beta|\boldsymbol{\hat{b}},\lambda)+g_A(\alpha|\boldsymbol{\hat{a}},\lambda)\,g_B(\beta|\boldsymbol{\hat{b}},\lambda).
\end{equation}  
A problem with the above expression is that it seems to trigger as immediate response its classification as ``nonlocal'', as a consequence of which it is taken to lie outside the range of local HV models. This comes from the unfortunate choice of the term ``nonlocal'' to describe something that is \emph{not} the logical negation of ``local'', as one would expect. This issue deserves to be discussed in more detail, as we do next.

\subsection{Locality and non-locality: two properties that do not exclude each other}

In the context of Bell inequalities, the term ``local'' has a precise meaning. It is technically used to mean that physical influences do not travel faster than light, which is also referred to as ``relativistic causality'' (see, e.g., \cite{hensen1,giustina,shalm} and references therein). In contrast to the precise meaning of ``local'', the term ``nonlocal'' has acquired a rather unprecise meaning in the context of the EPR paradox and the discussions that it prompted. Entanglement played a central role for the EPR paradox and ended up being considered a genuine quantum feature. QM was said to be a ``nonlocal'' theory, because it displays ``nonlocal correlations''. At the same time, ``Quantum mechanics, which does not allow us to transmit signals faster than light, preserves relativistic causality'' \cite{popescu1993}. QM is therefore both ``local'' and ``nonlocal'', a paradoxical sounding assertion, which is the consequence of an unfortunate choice of nomenclature. 

An expression such as the one in Eq.~(\ref{factor3}) may be dubbed nonlocal in the above sense, whenever one can derive from it nonlocal correlations. However, this does not imply that said expression must represent a non-classical situation. To see this, let us consider an everyday case. Suppose we know that Alice and Bob are a couple who owns a grocery store and who has agreed that, when one of them goes to their grocery store, the other stays home. We can describe our knowledge of Alice and Bob's case with the language of symbolic logic and write $(A_g \land B_h)\lor(A_h\land B_g)$, where $A_g$ means ``Alice is at the grocery store'', etc. Under the map $\land \leftrightarrow \otimes$ and $\lor \leftrightarrow +$, we can write the following correspondence:
\begin{equation}\label{correspondence}
(A_g \land B_h)\lor(A_h\land B_g) \longleftrightarrow |A\rangle_g\otimes |B\rangle_h + |A\rangle_h\otimes |B\rangle_g.
\end{equation}
Neither the logical expression nor its mathematical counterpart should be understood as implying by itself some physical ``delocalization'' of the involved entities. At any time, Alice and Bob are located at well defined places. However, all we can say about their locations is written down in the logical or else in the mathematical expression. If we see that Alice is at the grocery store, then ``we can predict with certainty (i.e., with probability equal to unity)'' that Bob is at home. This does not involve any physical ``collapse'', nor any action-at-a-distance. It is just EPR's ``criterion of reality'' \cite{epr}. Similar remarks apply to the expression for $p_{AB}(\alpha,\beta|\boldsymbol{\hat{a}},\boldsymbol{\hat{b}},\lambda)$ in Eq.~(\ref{factor3}).

It will be shown below that, on assuming a probability density having the structure of that in Eq.~(\ref{factor3}), we can reproduce the predictions of QM about probabilities and correlations entering the Bell-CHSH inequality. This being the case, the term ``nonlocal'' consistently applies to the $p_{AB}(\alpha,\beta|\boldsymbol{\hat{a}},\boldsymbol{\hat{b}},\lambda)$ of Eq.~(\ref{factor3}). However, this $p_{AB}(\alpha,\beta|\boldsymbol{\hat{a}},\boldsymbol{\hat{b}},\lambda)$ may be used in a \emph{local} realistic HV model, viz., a model which \emph{does not assume that there are physical influences that travel faster than light} and in which objects have properties that exist independently of being measured. Regarding entanglement, the expression for $p_{AB}(\alpha,\beta|\boldsymbol{\hat{a}},\boldsymbol{\hat{b}},\lambda)$ in Eq.~(\ref{factor3}) may be classified as an entangled one, unless it can be rewritten as a single product. This would be the case, if we go backwards after having applied a Lebesgue decomposition \cite{kolmogorov} to the probability density $p_{AB}(\alpha,\beta|\boldsymbol{\hat{a}},\boldsymbol{\hat{b}},\lambda)$ of Eq.~(\ref{factor}). Such a decomposition lets us write both
$p(\alpha|\boldsymbol{\hat{a}},\lambda)$ and $p(\beta|\boldsymbol{\hat{b}},\lambda)$ as the sum of three terms. That is, 
$p(\alpha|\boldsymbol{\hat{a}},\lambda)=p_1(\lambda)+p_2(\lambda)+p_3(\lambda)$, where $p_1(\lambda)$ is an absolutely continuous function, $p_2(\lambda)$ is a singular function and $p_3(\lambda)$ is a jump function. On applying a Lebesgue decomposition to  $p(\alpha|\boldsymbol{\hat{a}},\lambda)$ and $p(\beta|\boldsymbol{\hat{b}},\lambda)$ we get for
$p_{AB}(\alpha,\beta|\boldsymbol{\hat{a}},\boldsymbol{\hat{b}},\lambda)= p(\alpha|\boldsymbol{\hat{a}},\lambda)\,p(\beta|\boldsymbol{\hat{b}},\lambda)$ an expression that is similar to the one of Eq.~(\ref{factor3}). This shows that Eq.~(\ref{factor3}) should not be taken as one that violates locality, just because of its mathematical structure. Such a structure underlies the measure $S_{\text{CHSH}}$ itself, see Eq.~(\ref{bm1}). As we mentioned before, the structure of $S_{\text{CHSH}}$ derives from considering the function
\begin{equation}\label{f}
 f(\lambda)=\left(A_{\boldsymbol{\hat{a}}}(\lambda)+A_{\boldsymbol{\hat{a}^{\prime}}}(\lambda)\right)B_{\boldsymbol{\hat{b}}}(\lambda)+
(A_{\boldsymbol{\hat{a}}}(\lambda)-A_{\boldsymbol{\hat{a}^{\prime}}}(\lambda))B_{\boldsymbol{\hat{b}^{\prime}}}(\lambda).
\end{equation}
Regarding the involved locations at two distant sites, the function $f(\lambda)$ has the same mathematical structure as the one given in Eq.~(\ref{factor3}). One cannot disqualify one of these expressions for being at the basis of a local model or a locality test, without disqualifying the other as well. Entanglement can also be identified in Eqs.~(\ref{f}) and (\ref{factor3}), without thereby implying any connection with the quantum domain. It is worthwhile to mention here that physical realizations of entanglement with classical resources have been analyzed and reported by several authors \cite{spreeuw,eberly1,qian,borges,kagalwala,stoklasa,mclaren,sandeau}.

In summary, the probability density in Eq.~(\ref{factor}) cannot be considered to be the only one that complies with locality. For this reason, its factorized structure should be properly referred to as \emph{Bell locality hypothesis}, as was done, e.g., in Ref.~\cite{genovese2025}.
One may consider alternative expressions for $p(\alpha,\beta|\boldsymbol{\hat{a}},\boldsymbol{\hat{b}},\lambda)$, provided they comply with being non-negative and normalizable. Before addressing this task, we derive a direct consequence of Eq.~(\ref{factor3}). Upon integration, this equation leads to  
   
\begin{eqnarray}
P_{\boldsymbol{\hat{a}},\boldsymbol{\hat{b}}}(\alpha,\beta)&\equiv&\int_{\Lambda} p_{AB}(\alpha,\beta|\boldsymbol{\hat{a}},\boldsymbol{\hat{b}},\lambda)\rho(\lambda)d\lambda
 =\int_{\Lambda} f_A(\alpha|\boldsymbol{\hat{a}},\lambda)\,f_B(\beta|\boldsymbol{\hat{b}},\lambda)\rho(\lambda) d\lambda +
 \int_{\Lambda} g_A(\alpha|\boldsymbol{\hat{a}},\lambda)\,g_B(\beta|\boldsymbol{\hat{b}},\lambda)\rho(\lambda) d\lambda \nonumber \\
 &\equiv& P^{f}_{\boldsymbol{\hat{a}},\boldsymbol{\hat{b}}}(\alpha,\beta)+P^{g}_{\boldsymbol{\hat{a}},\boldsymbol{\hat{b}}}(\alpha,\beta).\label{connection2}
\end{eqnarray}
The above relation corresponds
to the probability rule $P(E)=P(E_f)+P(E_g)$ for an event $E$ occurring, which is the union of two events that exclude each other: $E=E_f \cup E_g$ with $E_f \cap E_g=\emptyset$. That is, if $P(E_f)\neq 0$, then $P(E_g)= 0$, and viceversa. We show next how a local realist model can be constructed, which is based on the decomposition given by Eq.~(\ref{factor3}).

\subsection{Proposed local realist model outside of Bell's framework}

The Bell-CHSH inequality has been violated in experiments performed with maximally entangled states, in particular with the singlet state
\begin{equation}\label{psi}
 |\psi^{-}\rangle=\frac{1}{\sqrt{2}}(|\uparrow\rangle|\downarrow\rangle-|\downarrow\rangle|\uparrow\rangle).
\end{equation}
For this state, QM yields 
\begin{eqnarray}\label{probqm}
  P^{\text{QM}}_{\psi^{-}}(\alpha,\beta|\boldsymbol{\hat{a}},\boldsymbol{\hat{b}})&=&\frac{1}{4}\left(1-\alpha\beta {\boldsymbol{\hat{a}}\cdot\boldsymbol{\hat{b}}} \right), \\
  \langle A B\rangle^{\text{QM}}_{\psi^{-}}&=&\sum_{\alpha,\beta}\alpha \beta P^{\text{QM}}_{\psi^{-}}(\alpha,\beta|\boldsymbol{\hat{a}},\boldsymbol{\hat{b}})=-{\boldsymbol{\hat{a}}\cdot\boldsymbol{\hat{b}}}.
\end{eqnarray}
Our task is then to choose $p_{AB}(\alpha,\beta|\boldsymbol{\hat{a}},\boldsymbol{\hat{b}},\lambda)$, so that
\begin{equation}\label{probhv}
  \int_{\Lambda} p_{AB}(\alpha,\beta|\boldsymbol{\hat{a}},\boldsymbol{\hat{b}},\lambda)\rho(\lambda)d\lambda=\frac{1}{4}\left(1-\alpha\beta {\boldsymbol{\hat{a}}\cdot\boldsymbol{\hat{b}}} \right).
\end{equation}
This task can be accomplished by extending to the two-qubit, singlet state the HV model of a single qubit, as given by Bell \cite{bell1964} or by Kochen and Specker \cite{kochen,rudolph}. In these models, the HVs are taken to be three-dimensional unit vectors $\boldsymbol{\hat{\lambda}}$ with a uniform distribution over the unit, 2D-sphere. Our model has a probability density with the structure that was displayed in Eq.~(\ref{factor3}) and reads 
\begin{equation}\label{ks}
  p_{AB}(\alpha,\beta|\boldsymbol{\hat{a}},\boldsymbol{\hat{b}},\boldsymbol{\hat{\lambda}})=(\alpha\, \boldsymbol{\hat{a}}\cdot \boldsymbol{\hat{\lambda}})\Theta(\alpha\, \boldsymbol{\hat{a}}\cdot \boldsymbol{\hat{\lambda}})\Theta(-\beta\, \boldsymbol{\hat{b}}\cdot \boldsymbol{\hat{\lambda}})+
  \Theta(-\alpha\, \boldsymbol{\hat{a}}\cdot \boldsymbol{\hat{\lambda}})(\beta\, \boldsymbol{\hat{b}}\cdot \boldsymbol{\hat{\lambda}})\Theta(\beta\, \boldsymbol{\hat{b}}\cdot \boldsymbol{\hat{\lambda}}),
\end{equation}
where $\Theta (x)$ is the step function: $\Theta (x)=1$ for $x\geq 0$
and $\Theta (x)=0$ for $x<0$. As for the probability density of the uniformly distributed HVs, we set $\rho(\boldsymbol{\hat{\lambda}})=1/4\pi$. Our HVs are thus the polar and azimuthal angles of the unit vectors $\boldsymbol{\hat{\lambda}}=(\sin \theta _{\lambda }\cos \varphi _{\lambda },\sin \theta_{\lambda }\sin \varphi _{\lambda },\cos \theta _{\lambda })$, with $\theta _{\lambda }\in [0,\pi]$ and $\varphi _{\lambda }\in[0,2\pi]$ defining the domain $\Lambda$, while $d\boldsymbol{\hat{\lambda}}=\sin \theta _{\lambda }d\theta_{\lambda }d\varphi _{\lambda }$, so that $\int_{\Lambda} \rho(\boldsymbol{\hat{\lambda}})d\boldsymbol{\hat{\lambda}}=1$.

To prove that Eq.~(\ref{ks}) gives Eq.~(\ref{probhv}), it suffices to show that the following result holds true:
\begin{equation}
I_{ab}=
\int_{\Lambda} (\boldsymbol{\hat{a}}\cdot \boldsymbol{\hat{\lambda}}) \Theta(\boldsymbol{\hat{a}}\cdot \boldsymbol{\hat{\lambda}}) \Theta(\boldsymbol{\hat{b}}\cdot \boldsymbol{\hat{\lambda}})d\boldsymbol{\hat{\lambda}} =\frac{\pi}{2}\left( 1+\boldsymbol{\hat{a}}\cdot
\boldsymbol{\hat{b}}\right).  \label{iab}
\end{equation}
This equation follows from first observing that $I_{ab}=\int_{S_{ab}}( \boldsymbol{\hat{a}}\cdot
\boldsymbol{\hat{\lambda}}) d\boldsymbol{\hat{\lambda}}$, where $S_{ab}$ is the intersection of the
two northern hemispheres that are defined by $\boldsymbol{\hat{a}}$ and $\boldsymbol{\hat{b}}$. Hence, the boundary of $S_{ab}$ consists of two half great circles, $C_{a}$ and $C_{b}$.
The flux integral  $\int_{S_{ab}}( \boldsymbol{\hat{a}}\cdot
\boldsymbol{\hat{\lambda}}) d\boldsymbol{\hat{\lambda}}$ can be calculated with the
help of Stokes theorem. Indeed, let us define the vector field $\boldsymbol{v}_{a}(\boldsymbol{r})=\boldsymbol{\hat{a}}\times \boldsymbol{r}/2$.
We have then that $\boldsymbol{\hat{a}}=\nabla \times \boldsymbol{v}_{a}$, so that $I_{ab}=\int_{S_{ab}}\left( \nabla \times \boldsymbol{v}_{a}\right)\cdot \boldsymbol{\hat\lambda}\, 
d\boldsymbol{\hat{\lambda}} =\oint\limits_{\partial S_{ab}}\boldsymbol{v}_{a}(\boldsymbol{\hat{r}}_{\lambda})\cdot
d\boldsymbol{\hat{r}}_{\lambda} =(1/ 2)\oint\limits_{\partial S_{ab}}(\boldsymbol{\hat{a}}\times
\boldsymbol{\hat{r}}_{\lambda})\cdot d\boldsymbol{\hat{r}}_{\lambda} =(1/ 2)\oint\limits_{\partial
S_{ab}}( \boldsymbol{\hat{r}}_{\lambda} \times d\boldsymbol{\hat{r}}_{\lambda} /ds)\cdot \boldsymbol{\hat{a}}\, ds$. Here, 
$\partial S_{ab}$ is the boundary of $S_{ab}$ and $s$ is the arc-length used to parameterize the curve $\boldsymbol{\hat{r}}_{\lambda}(s)=\boldsymbol{\hat{\lambda}}(s)$ on the sphere. Hence, $d\boldsymbol{\hat{r}}_{\lambda} /ds$ is a unit
vector, tangent to the sphere. On noting that $\partial S_{ab}=C_{a}\cup C_{b}$
and that $\boldsymbol{\hat{r}_{\lambda}} \times (d\boldsymbol{\hat{r}}_{\lambda} /ds)$ is a unit vector that equals $\boldsymbol{\hat{a}}$ along $C_{a}$ and $\boldsymbol{\hat{b}}$ along $C_{b}$, we see that
\begin{equation}
I_{ab}=\frac{1}{2}\left[ \int_{C_{a}}\left( \boldsymbol{\hat{a}}\cdot \boldsymbol{\hat{a}}\right)
ds+\int_{C_{b}}( \boldsymbol{\hat{b}}\cdot \boldsymbol{\hat{a}}) ds\right] =\frac{\pi}{2}\left( 1+\boldsymbol{\hat{a}}\cdot
\boldsymbol{\hat{b}}\right) .  \label{1c}
\end{equation}
We can use this result and Eq.~(\ref{ks}) to calculate
\begin{eqnarray}
   P^{\text{HV}}_{\psi^{-}}(\alpha,\beta|\boldsymbol{\hat{a}},\boldsymbol{\hat{b}})&\equiv&\int_{\Lambda} p_{AB}(\alpha,\beta|\boldsymbol{\hat{a}},\boldsymbol{\hat{b}},\boldsymbol{\hat{\lambda}})\rho(\boldsymbol{\hat{\lambda}})d\boldsymbol{\hat{\lambda}} \label{probhv2a}\\
   &=&
  \frac{1}{4\pi}\int_{\Lambda} \left[(\alpha\, \boldsymbol{\hat{a}}\cdot \boldsymbol{\hat{\lambda}})\Theta(\alpha\, \boldsymbol{\hat{a}}\cdot \boldsymbol{\hat{\lambda}})\Theta(-\beta\, \boldsymbol{\hat{b}}\cdot \boldsymbol{\hat{\lambda}})+
  \Theta(-\alpha\, \boldsymbol{\hat{a}}\cdot \boldsymbol{\hat{\lambda}})(\beta\, \boldsymbol{\hat{b}}\cdot \boldsymbol{\hat{\lambda}})\Theta(\beta\, \boldsymbol{\hat{b}}\cdot \boldsymbol{\hat{\lambda}})\right]d\boldsymbol{\hat{\lambda}} \label{probhv2b}\\
  &=&\frac{1}{4\pi}\left[\frac{\pi}{2}(1+ ( {\alpha\boldsymbol{\hat{a}})\cdot(-\beta\boldsymbol{\hat{b}}}))+\frac{\pi}{2}(1+(-\alpha{\boldsymbol{\hat{a}})(\beta \cdot\boldsymbol{\hat{b}}})) \right] =\frac{1}{4}\left(1-\alpha\beta {\boldsymbol{\hat{a}}\cdot\boldsymbol{\hat{b}}} \right) \label{probhv2d}.
\end{eqnarray}
$P^{\text{HV}}_{\psi^{-}}(\alpha,\beta|\boldsymbol{\hat{a}},\boldsymbol{\hat{b}})$ is in accordance with Eq.~(\ref{connection2}), the probability of two mutually exclusive events. The two summands in the first equality of  Eq.~(\ref{probhv2d}) stem from mutually exclusive subsets of the integration domain $\Lambda$. The final result in Eq.~(\ref{probhv2d}) merges the two contributions into a single expression, which does not allow to distinguish one contribution from the other. This final form coincides with the quantum result. Having obtained $P^{\text{HV}}_{\psi^{-}}(\alpha,\beta|\boldsymbol{\hat{a}},\boldsymbol{\hat{b}})=P^{\text{QM}}_{\psi^{-}}(\alpha,\beta|\boldsymbol{\hat{a}},\boldsymbol{\hat{b}})$,
and hence
\begin{equation}\label{hvcorr}
  \langle A B\rangle^{HV}_{\psi^{-}}=\sum_{\alpha,\beta}\alpha \beta P^{\text{HV}}_{\psi^{-}}(\alpha,\beta|\boldsymbol{\hat{a}},\boldsymbol{\hat{b}})=-{\boldsymbol{\hat{a}}\cdot\boldsymbol{\hat{b}}}=\langle A B\rangle^{QM}_{\psi^{-}},
\end{equation}
our HV model is in agreement with experimental outputs, in particular with those that violate the Bell-CHSH inequality. 

The singlet state $|\psi^{-}\rangle$ we have addressed is one of the four Bell states: $|\phi^{\pm}\rangle = (|\uparrow\rangle|\uparrow\rangle\pm|\downarrow\rangle|\downarrow\rangle)/\sqrt{2}$, 
$|\psi^{\pm}\rangle = (|\uparrow\rangle|\downarrow\rangle\pm|\downarrow\rangle|\uparrow\rangle)/\sqrt{2}$. Any of these states can be used to produce violations of the Bell-CHSH inequality. Besides $\langle A B\rangle^{QM}_{\psi^{-}}=-a_1 b_1-a_2 b_2-a_3 b_3=-{\boldsymbol{\hat{a}}\cdot\boldsymbol{\hat{b}}}$, the other quantum correlations are given by $\langle A B\rangle^{QM}_{\psi^{+}}=a_1 b_1+a_2 b_2-a_3 b_3$, $\langle A B\rangle^{QM}_{\phi^{-}}=-a_1 b_1 +a_2 b_2+a_3 b_3$, and
$\langle A B\rangle^{QM}_{\phi^{+}}=a_1 b_1-a_2 b_2 +a_3 b_3$. A HV model can be constructed for a Bell-CHSH test that uses any of the Bell states. We just need to appropriately define one of the unit vectors entering the model, say $\boldsymbol{\hat{a}}$. For example, if a Bell-CHSH test uses $|\phi^{+}\rangle$, then we define our HV model to be the same as that of $|\psi^{-}\rangle$, but with $\boldsymbol{\hat{a}}=(a_1,-a_2,a_3)$. We can proceed similarly with the other Bell states.

\subsection{Comparison with other models that reproduce quantum predictions}

It is worthwhile to stress that, whereas the probability density $ p_{AB}(\alpha,\beta|\boldsymbol{\hat{a}},\boldsymbol{\hat{b}},\lambda)$ in Eq.~(\ref{factor3}) generally does not have Bell's factorable form, see Eq.~(\ref{factor}), its particular implementation given by Eq.~(\ref{ks}) effectively does not differ from Bell's form. Indeed, for each $\boldsymbol{\hat{\lambda}}$, at most one of the two summands entering Eq.~(\ref{ks}) is not zero and it has Bell's factorable form. Even so, our HV model lies outside the domain of Bell's theorem. This is also the case with other models, to which we briefly refer here, for the sake of comparison. In what follows, the assumptions underlying Bell's theorem, which we listed at the beginning, will be reformulated somewhat, following Ref. \cite{vieira2025}. This helps to identify which assumptions are not satisfied by the HV models that are outside the domain of Bell's theorem.

The fundamental assumption in Bell's theorem is given by the first equality in Eq.~(\ref{connection1}), i.e.

\begin{equation}\label{a1}
 P(\alpha,\beta|\boldsymbol{\hat{a}},\boldsymbol{\hat{b}})\equiv P_{\boldsymbol{\hat{a}},\boldsymbol{\hat{b}}}(\alpha,\beta)=\int_{\Lambda} p(\alpha,\beta|\boldsymbol{\hat{a}},\boldsymbol{\hat{b}},\lambda)\rho(\lambda|\boldsymbol{\hat{a}},\boldsymbol{\hat{b}}) d\lambda.
\end{equation}
Here, we have replaced the HVs density $\rho(\lambda)$ by $\rho(\lambda|\boldsymbol{\hat{a}},\boldsymbol{\hat{b}})$. This leaves open the option that HVs and measurement settings are correlated. Upon the assumption that HVs have no influence on the settings, i.e., $p(\boldsymbol{\hat{a}},\boldsymbol{\hat{b}}|\lambda)=p(\boldsymbol{\hat{a}},\boldsymbol{\hat{b}})$, Bayes' theorem, $P(A|B)=P(B|A)P(A)/P(B)$, leads to
\begin{equation}\label{a2}
\rho(\lambda|\boldsymbol{\hat{a}},\boldsymbol{\hat{b}})=
\frac{p(\boldsymbol{\hat{a}},\boldsymbol{\hat{b}}|\lambda)\rho(\lambda)}{p(\boldsymbol{\hat{a}},\boldsymbol{\hat{b}})}=\rho(\lambda).
\end{equation}
This result, $\rho(\lambda|\boldsymbol{\hat{a}},\boldsymbol{\hat{b}})=\rho(\lambda)$, is called \emph{measurement independence} (MI). Two additional assumptions are \emph{outcome independence}
\begin{equation}\label{a3}
  \text{OI:}\quad p(\alpha|\boldsymbol{\hat{a}},\boldsymbol{\hat{b}},\beta,\lambda)=p(\alpha|\boldsymbol{\hat{a}},\boldsymbol{\hat{b}},\lambda), \quad
  p(\beta|\boldsymbol{\hat{a}},\boldsymbol{\hat{b}},\alpha,\lambda)=p(\beta|\boldsymbol{\hat{a}},\boldsymbol{\hat{b}},\lambda),
\end{equation}
and \emph{parameter independence}
\begin{equation}\label{a4}
  \text{PI}:\quad
  p(\alpha|\boldsymbol{\hat{a}},\boldsymbol{\hat{b}},\lambda)=p(\alpha|\boldsymbol{\hat{a}},\lambda), \quad
  p(\beta|\boldsymbol{\hat{a}},\boldsymbol{\hat{b}},\lambda)=p(\beta|\boldsymbol{\hat{b}},\lambda).
\end{equation}
OI and PI together imply Bell's factorability \cite{vieira2025}. Indeed, from the definition of conditional probability, $P(A|B)=P(A,B)/P(B)$, we get 
$p(\beta|\boldsymbol{\hat{a}},\boldsymbol{\hat{b}},\lambda)=p(\boldsymbol{\hat{a}},\boldsymbol{\hat{b}},\beta,\lambda)/p(\boldsymbol{\hat{a}},\boldsymbol{\hat{b}},\lambda)$.
On using this result, we get
\begin{equation}\label{a5}
\frac{p(\alpha,\beta|\boldsymbol{\hat{a}},\boldsymbol{\hat{b}},\lambda)}{p(\beta|\boldsymbol{\hat{a}},\boldsymbol{\hat{b}},\lambda)}=
\frac{p(\alpha,\beta|\boldsymbol{\hat{a}},\boldsymbol{\hat{b}},\lambda)p(\boldsymbol{\hat{a}},\boldsymbol{\hat{b}},\lambda)}
{p(\boldsymbol{\hat{a}},\boldsymbol{\hat{b}},\beta,\lambda)}=\frac{p(\alpha,\beta,\boldsymbol{\hat{a}},\boldsymbol{\hat{b}},\lambda)}
{p(\boldsymbol{\hat{a}},\boldsymbol{\hat{b}},\beta,\lambda)}=p(\alpha|\boldsymbol{\hat{a}},\boldsymbol{\hat{b}},\beta,\lambda),
\end{equation}
from which it follows that
\begin{equation}\label{a6}
p(\alpha,\beta|\boldsymbol{\hat{a}},\boldsymbol{\hat{b}},\lambda)=p(\alpha|\boldsymbol{\hat{a}},\boldsymbol{\hat{b}},\lambda,\beta)p(\beta|\boldsymbol{\hat{a}},\boldsymbol{\hat{b}},\lambda).
\end{equation}
On invoking first OI and then PI, we get $p(\alpha|\boldsymbol{\hat{a}},\boldsymbol{\hat{b}},\lambda,\beta)=p(\alpha|\boldsymbol{\hat{a}},\boldsymbol{\hat{b}},\lambda)=p(\alpha|\boldsymbol{\hat{a}},\lambda)$.
Moreover, because of PI, $p(\beta|\boldsymbol{\hat{a}},\boldsymbol{\hat{b}},\lambda)=p(\beta|\boldsymbol{\hat{b}},\lambda)$. Replacing these two results in Eq.~(\ref{a6}), we get
\begin{equation}\label{a7}
p(\alpha,\beta|\boldsymbol{\hat{a}},\boldsymbol{\hat{b}},\lambda)=
p(\alpha|\boldsymbol{\hat{a}},\lambda)\,p(\beta|\boldsymbol{\hat{b}},\lambda),
\end{equation}
which is Bell's factorability.

It should be stressed that Bayes' theorem and the definition of conditional probabilities have been used to get MI and Bell's factorability, Eqs.~(\ref{a2}) and (\ref{a7}), respectively. That is, it has been assumed that what holds for probabilities also holds for probability densities. This is questionable. Indeed, on assuming that $p(A,B,\lambda)=p(A|B,\lambda)p(B,\lambda)$, upon integration we obtain
\begin{equation}\label{a8}
P(A,B)=\int p(A,B,\lambda)d\lambda=\int p(A|B,\lambda)p(B,\lambda)d\lambda.
\end{equation}
Generally, the last integral does not factorize to give $P(A|B)P(B)$, as it is required for $P(A,B)$ to be in accordance with the definition of conditional probability: $P(A|B)=P(A,B)/P(B)$. We have here the same state of affairs that we had before, when dealing with Bell's factorability applied to probability densities instead of probabilities. We are thus led to ask the following question, which is pertinent to the general discussion of HV models:

\medskip

Can we require that a probability density, say $p(\alpha,\beta|\boldsymbol{\hat{a}},\boldsymbol{\hat{b}},\lambda)$, satisfies physical requirements that are \emph{not} satisfied by the corresponding probability $P(\alpha,\beta|\boldsymbol{\hat{a}},\boldsymbol{\hat{b}})$?  

\medskip

We believe that the answer to this question should be in the negative, because what we submit to experimental test is a probability, not a probability density. However, the discussions of HV models often refer to probability densities as if they were probabilities. Bell locality assumption is an example of this, as we have seen.

Coming back to the assumptions underlying Bell's theorem, on relaxing one or more of them, one can achieve accordance with quantum predictions. For example, MI can be relaxed by assuming that HVs in the remote past fully determine the choice of measurement settings \cite{brans1988}, thereby negating ``free will''. Alternatively, one can define some specific correlation between HVs and measurement settings \cite{dilorenzo2012b}. In the two cases, one reproduces the quantum predictions for the singlet state. Also OI and/or PI can be relaxed to assume some degree of non-locality or action-at-a-distance. The so-called ``transactional interpretation'' of QM \cite{cramer1986} is based on such an assumption, which would explain Bell inequality violations. The assumed, underlying physical mechanism is the exchange of advanced and retarded waves, which is in turn the basis of the Wheeler-Feynman ``absorber theory'' of electrodynamics \cite{wheeler1949}.   

As compared to the above departures from Bell's framework, ours seems to be a rather mild one. The HV model we have presented is a local and realist one that provides a non-quantum description of all reported outcomes of the Bell-CHSH experiment. As noticed before, the probability density $p_{AB}(\alpha,\beta|\boldsymbol{\hat{a}},\boldsymbol{\hat{b}},\boldsymbol{\hat{\lambda}})$ that is defined in Eq.~(\ref{ks}) is, for each $\boldsymbol{\hat{\lambda}}$, in Bell's factorable form, because at most only one of its two summands is different from zero. This fact by itself implies that none of the assumptions of Bell's theorem, MI, OI and PI, can be effectively in conflict with such a probability density. But even if the two summands in Eqs.~(\ref{ks}) or (\ref{factor3}) were different from zero at the same time, the probability density would satisfy MI, OI and PI, despite not being of factorable form. As we have seen, OI and PI imply Bell's factorability only under the assumption that the definition of conditional probabilities applies for probability densities. Such an assumption leads to an inconsistency. The obvious conclusion that follows from all this is that Bell's model is not the only one that complies with MI, OI and PI. Bell-CHSH inequality violations falsify Bell's particular model, but not all possible, 
local realist models.

\section{Closing remarks}

As we have seen, it is legitimate to consider a non-factorable probability density, irrespective of how we interpret it regarding non-locality. This probability density can be used to construct a HVs model that reproduces quantum predictions about the output of a Bell-CHSH test. At the very least, this HVs model represents a description of the Bell-CHSH experiment that is completely different from the quantum description. This makes clear that the Bell locality assumption, i.e. the factorizability of probability densities, is a too narrowly framed assumption to be set at the basis of all HVs models. It is an open question how far one can go in the construction of HVs models without such a restricting assumption. 

The mere existence of our HVs model implies that Bell-CHSH inequality violations cannot be considered as an authorized signature of quantumness, let alone a certification of the availability of entangled quantum states. Along with coherence, entanglement is considered a main resource for quantum technologies. A so-called second quantum revolution has been announced to take place when genuine quantum resources such as coherence and entanglement can be fully exploited. For example, quantum secure communication should be based on quantum key distribution, which in turn relies on entangled photon pairs shared by senders and recipients. As we have seen, it is possible to replicate quantum results without resorting to the quantum formalism. This begs the question of whether some quantum algorithms could be implemented with conventional technology. A positive answer to this question would allow reaching quantum advantage, even without having access to high-tech resources. 

Finally, it should be stressed that our results should not be misinterpreted as a refutation of Bell's theorem. The Bell-CHSH inequality is an indisputable logical consequence of three hypotheses, the third one being Bell's factorability of the \emph{probability density} that belongs to the joint \emph{probability} of getting measurement results $\alpha=\pm 1$ and $\beta=\pm 1$ for the respective settings $\boldsymbol{\hat{a}}$ and $\boldsymbol{\hat{b}}$. What we have shown is that, on replacing the third hypothesis by another one, quantum predictions can be replicated within a HVs model.

\end{document}